\documentclass[12pt,onecolumn,draftclsnofoot]{IEEEtran}

\usepackage{graphicx}
\usepackage{times}
\usepackage{amsmath}
\usepackage{color}
\usepackage{mathptm}
\usepackage{epsfig}
\usepackage{array}
\usepackage{epsfig}
\usepackage{slashbox}
\usepackage{diagbox}

\ifCLASSINFOpdf

\else

\fi

\begin{document}
%
\title{Optimal Distributed Control for Networked Control Systems with Delays}

\author{Zhuwei~Wang,
        and~Xiaodong Wang,~\IEEEmembership{Fellow,~IEEE,}
\thanks{Z. Wang and X. Wang are with the Electrical Engineering Department, Columbia University,
New York, 10027 (e-mail: wzw902@gmail.com, wangx@ee.columbia.edu).}
}


\maketitle

\begin{abstract}
In networked control systems (NCS), sensing and control signals between the plant and controllers
are typically transmitted wirelessly.
Thus, the time delay plays an important role for the stability of NCS, especially with distributed controllers.
In this paper, the optimal control strategy is derived for distributed control networks with time delays.
In particular, we form the optimal control problem as a non-cooperative linear quadratic game (LQG).
Then, the optimal control strategy of each controller is obtained that is based
on the current state and the last control strategies.
The proposed optimal distributed controller reduces to some known controllers under certain conditions.
Moreover, we illustrate the application of the proposed distributed controller to load frequency control in
power grid systems.
\end{abstract}


\IEEEpeerreviewmaketitle

\section{Introduction}
\IEEEPARstart{I}{n} recent years, networked control systems (NCS), which consist of computing and
physical systems, have received considerable attention \cite{IRef1} due to their wide
applications in various areas such as power grids \cite{IRef2}, robotic networks \cite{IRef3} and
embedded systems \cite{IRef4}. A typical NCS is equipped with sensing, control and communication capabilities.
In many cases, the plant and controllers are at different locations.
Hence, a communication network, typically a wireless network,
is needed to facilitate the data exchange between the plant and controllers.
Then the time delay becomes the key factor that affects the system performance and stability.

Existing works on NCS with time delay focus on the single-controller case;
and two important design considerations are the system stability and the optimality with respect to
certain criterion. With full plant state information, the optimal control problem has been investigated.
In particular, a suboptimal controller is derived in \cite{IRef5} with time-driven sensor and controller nodes,
where the time delay is a multiple of the sampling interval.
The optimal controller for an NCS whose network-induced delay is shorter than a sampling period
is developed in \cite{IRef6}\cite{IRef7}. And the results are generalized in \cite{IRef8}
to the case that network-induced delay is longer than a sampling period.
In \cite{IRef9}, the solution to the optimal control problem for a linear system with multiple
control input delays is given.
On the other hand, when considering the packet loss, partial state information, link/node failure, etc.,
only system stability can be investigated \cite{IRef10}-\cite{IRef14},
since the optimality problem is extremely difficult.

With the advances of NCS, the concept of distributed controllers in large scale systems becomes
an important research topic \cite{IRef15}-\cite{IRef17}.
A cross-layer framework for the joint design of wireless networks and distributed controllers is proposed
in \cite{IRef15}, where the centralized control and clock-driven controllers are considered and
the total time delay is assumed to be one sample period.
The stability of a distributed control strategy is studies in \cite{IRef16},
where the network itself acts as a controller, and each node (including the actuator nodes)
performs linear combinations of internal state variables of neighboring nodes.
The stability of a multicast routing algorithm for a decentralized control system is investigated
in \cite{IRef17} assuming no time delay or extremely small delay.
Note that the above works all address stability issues of distributed control,
but the optimality problem remains unexplored.

This paper addresses the optimal control problem for a linear distributed control system with time delays.
The form of the performance criterion plays an important role in obtaining the optimal solution:
previous studies have mostly focused on the quadratic cost function \cite{IRef5}-\cite{IRef9},
which is also used in this work.
In this paper, the optimal solution is obtained as a feedback non-cooperative control law,
which is linear with the current state and the previous control strategies of the distributed controllers.
An application of the proposed optimal distributed controller to load frequency control
in power grid is also described.

The remainder of this paper is organized as follows.
The system model and problem formulation are given in Section II.
We then derive the optimal control strategy with two distributed controllers in Section III.
Section IV presents the extension to the case of multiple distributed controllers.
Numerical results and conclusions are given in Section IV and Section V, respectively.

\section{System Model and Problem Formulation}
In this section, we first describe the distributed control system under consideration,
and then formulate the optimal control problem as a non-cooperative linear quadratic game (LQG).

\subsection{Distributed Control System}

\begin{figure}[h]
  \begin{center}
  \includegraphics[width=0.8\textwidth]{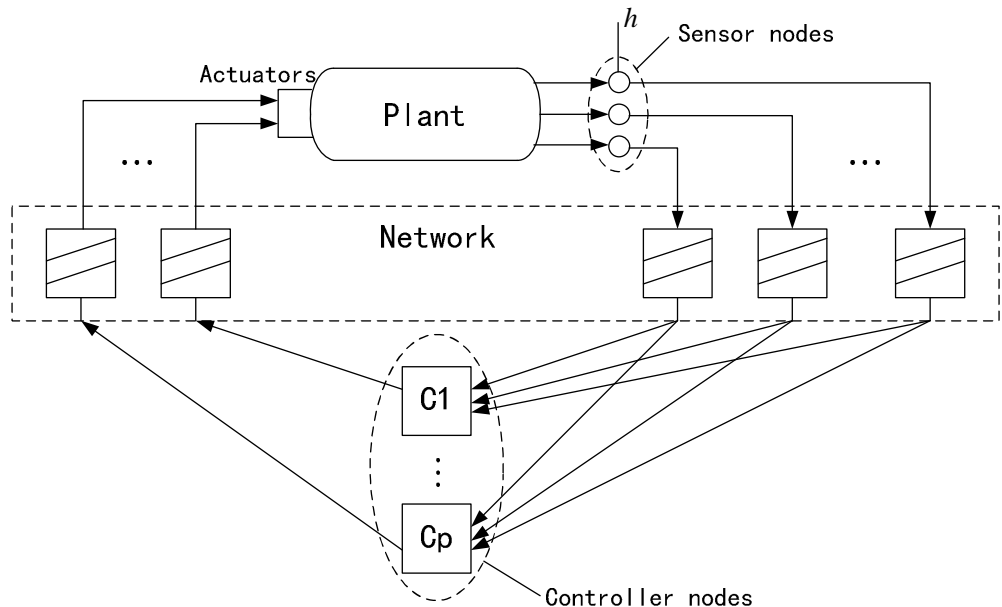}
  \caption{The structure of a networked distributed control system.} \label{Fig1}
  \end{center}
\end{figure}

We consider a networked control system with distributed sensors and controllers as shown in Fig.~\ref{Fig1}.
We assume that the plant is a continuous-time linear time-invariant (LTI) system while all sensors
and controllers operate in discrete-time. Sensor measurements and feedback control signals are
sent separately through a shared wireless network. The system under consideration has a
time-driven sensor system sampled at a constant sampling rate and event-driven controllers
and actuator nodes. We assume that there are $M$ sensor nodes and $p$ distributed controller nodes.
Then, free of perturbations, the continuous-time state and measurement equations are given respectively by
\begin{equation}\label{eq:1}
\begin{cases}
   {\dot x\left( t \right) = A^c x\left( t \right) + \sum\limits_{i = 1}^p {B_i^c u_i \left( {t - \tau _i } \right)} ,}  \\
   {y\left( t \right) = Cx\left( t \right),}  \\
\end{cases}
\end{equation}
where $x$ is an $M$-dimensional plant state vector,
$u_i$ is an $N$-dimensional $i$-$th$ control input vector and ${\tau _i }$ is the time delay,
$A^c$ and $B_i^c$ are $M \times M$ and $M \times N$ matrices, respectively.
For simplicity, we assume that each sensor observes one dimension of $x$ directly
so that $y$ is an $M$-dimensional vector and $C$ is an identity matrix.

\subsection{Problem Formulation}

\begin{figure}[h]
  \begin{center}
  \includegraphics[width=0.75\textwidth]{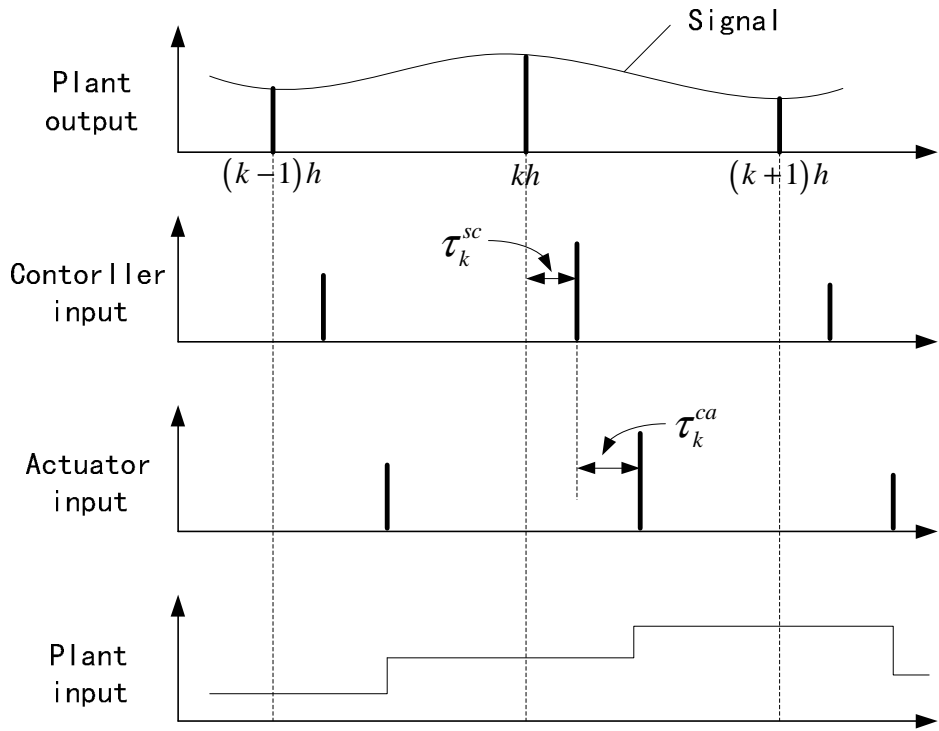}
  \caption{Timing of signals in the control system.} \label{Fig2}
  \end{center}
\end{figure}

We assume that there is a wireless communication network from sensors to controllers,
and sampling in the sensor nodes are done synchronously with the period $h$.
Upon sampling the measurements are immediately sent to the controller nodes.
Thus, the $M$ measurement signals will have individual
delays $\tau _{i,j}^{sc},\ {\rm{ }}j \in \left\{ {1,\ {\rm{ }}2,{\rm{ }} \cdots ,\ {\rm{ }}M} \right\}$,
to the $i$-$th$ controller node. When all measurements have arrived at the controller,
a new control signal is calculated and sent to the actuator nodes.
The time delay from sensors to the $i$-$th$ controller node is
$\tau _i^{sc}  = \max \left\{ {\tau _{i,1}^{sc} ,\ {\rm{ }}\tau _{i,2}^{sc} ,{\rm{ }} \cdots {\rm{, }}\ \tau _{i,M}^{sc} } \right\}$,
and the control signals will have delays
$\tau _i^{ca} ,\ {\rm{ }}i = \left\{ {1,\ {\rm{ }}2,{\rm{ }} \cdots ,\ {\rm{ }}p} \right\}$,
to the actuator nodes. We assume that all delays in the system are deterministic and known,
and as in \cite{IRef6}\cite{IRef7} the total time
delay $\tau _i^{sc}  + \tau _i^{ca}$ is assumed to be smaller than one sampling period.
The received control signals are then converted to continuous-time signals and directly act on the plant.
The signal timings in the control system are illustrated in Fig.~\ref{Fig2}.

We assume that all nodes have synchronized clocks. This is needed both for synchronized sampling and
time-stamping of signals. By the use of time-stamping we assume that all delays are known to the controller node.
Then, discretizing the process in (\ref{eq:1}) gives
\begin{equation}\label{eq:2}
x\left( {k + 1} \right) = \Phi x\left( k \right) + \sum\limits_{i = 1}^p {\left[ {\Gamma _{i,0} u_i \left( k \right) + \Gamma _{i,1} u_i \left( k-1 \right)} \right],}
\end{equation}
where $x\left( k \right)$ and $u_i \left( k \right)$ represent the state and the control signals at
the $k$-$th$ sampling instant, respectively, and
\begin{equation}\label{eq:3}
\begin{split}
 \Phi  &= e^{A^c h} , \\
 \Gamma _{i,0}  &= \int_0^{h - \tau _i^{sc}  - \tau _i^{ca} } {e^{A^c s} ds} B_i^c , \\
 \Gamma _{i,1}  &= \int_{h - \tau _i^{sc}  - \tau _i^{ca} }^h {e^{A^c s} ds} B_i^c . \\
 \end{split}
\end{equation}

Using a quadratic cost function, the design problem is to find a control strategy to drive the plant
from its initial state $x_0$ to minimize the total cost, i.e.,
\begin{equation}\label{eq:4}
\begin{split}
 \mathop {\min }\limits_{\scriptstyle u_i \left( k \right),{\rm{ }}\ k = 0,\ 1, \cdots ,\ N - 1 \hfill \atop
  \scriptstyle {\rm{          }}\quad \ \ \  i = 1,\ 2, \cdots ,\ p \hfill} {\rm{  }}\ \ &J_N  = x^T \left( N \right)Q_N x\left( N \right) + \sum\limits_{k = 0}^{N - 1} {\left\{ {x^T \left( k \right)Qx\left( k \right) + \sum\limits_{i = 1}^p {u_i^T \left( k \right)R_i u_i \left( k \right)} } \right\}},  \\
 {\rm{           }}s.t.\ \ &{\rm{   }}x\left( {k + 1} \right) = \Phi x\left( k \right) + \sum\limits_{i = 1}^p {\left[ {\Gamma _{i,0} u_i \left( k \right) + \Gamma _{i,1} u_i \left( k - 1 \right)} \right]},  \\
 &{\rm{                  }}x\left( 0 \right) = x_0,  \\
 \end{split}
\end{equation}
where $N$ is the total number of sampling instants,
$Q_N  \succeq 0$, $Q \succ 0$, and $R_i \succ 0$ are symmetric positive semi-definite/definite weight matrices.

Since the controllers are distributed and they cannot obtain the current control strategies of each other,
we reformulate the optimization problem in (\ref{eq:4}) as a non-cooperative control game as \cite{IRef18}
\begin{equation}\label{eq:5}
\begin{split}
 \mathop {\min }\limits_{u_i \left( k \right),\ {\rm{ }}k = 0,\ 1, \cdots ,\ N - 1} {\rm{  }}\ \ &J_{i,N}  = x^T \left( N \right)Q_{i,N} x\left( N \right) + \sum\limits_{k = 0}^{N - 1} {\left\{ {x^T \left( k \right)Q_i x\left( k \right) + u_i^T \left( k \right)R_i u_i \left( k \right)} \right\}} ,\ \ {\rm{ }}\forall i ,\\
 {\rm{           }}s.t.\ \ &{\rm{   }}x\left( {k + 1} \right) = \Phi x\left( k \right) + \sum\limits_{j = 1}^p {\left[ {\Gamma _{j,0} u_j \left( k \right) + \Gamma _{j,1} u_j \left( k - 1 \right)} \right]},  \\
 &{\rm{                  }}x\left( 0 \right) = x_0,  \\
 \end{split}
\end{equation}
where $Q_{i,N} \succeq 0$ and $Q_i\succ 0$ are symmetric weight matrices.

In what follows, we first focus on a two-controller distributed system, i.e., $p=2$,
and then extend the results to the case with multiple distributed controllers.

\section{Optimal Solution for Two-Controller Case}

In this section, we fist derive the optimal linear control strategy for the non-cooperative game in (\ref{eq:5})
with two distributed controllers, and then consider two special cases, where the obtained optimal controller
becomes some existing controllers in the literature.

\subsection{Derivation of Optimal Controllers}

Since the two controllers are distributed, at any time,
the current control signal of one controller is not known to the other.
We assume that each controller can obtain the other controller's past control signals.
Then, the linear control law can be written as
\begin{equation}\label{eq:6}
u_i \left( k \right) = A_i \left( k \right)x\left( k \right) + \sum\limits_{j = 1}^k {\left[ {B_{1,j}^i \left( k \right)u_1 \left( {k - j} \right) + B_{2,j}^i \left( k \right)u_2 \left( {k - j} \right)} \right]} ,\ i = 1,\ 2,
\end{equation}
where $A_i \left( k \right),\ B_{1,j}^i \left( k \right),\ B_{2,j}^i \left( k \right)$ are $N \times M$, $N \times N$, $N \times N$ coefficient matrices, respectively.

Taking controller 1 as the desired controller,
substituting $u_2 \left( k \right)$ in (\ref{eq:6}) into (\ref{eq:5}), we have
\begin{equation}\label{eq:7}
\begin{split}
 x\left( {k + 1} \right) = &\left[ {\Phi  + \Gamma _{2,0} A_2 \left( k \right)} \right]x\left( k \right) + \left[ {\Gamma _{1,1}  + \Gamma _{2,0} B_{1,1}^2 \left( k \right)} \right]u_1 \left( {k - 1} \right) + \left[ {\Gamma _{2,1}  + \Gamma _{2,0} B_{2,1}^2 \left( k \right)} \right]u_2 \left( {k - 1} \right) \\
 {\rm{                }} &+ \Gamma _{1,0} u_1 \left( k \right) + \sum\limits_{i = 2}^k {\left[ {\Gamma _{2,0} B_{1,i}^2 \left( k \right)u_1 \left( {k - i} \right) + \Gamma _{2,0} B_{2,i}^2 \left( k \right)u_2 \left( {k - i} \right)} \right]}.  \\
 \end{split}
\end{equation}

Define
\begin{equation}\label{eq:8}
z\left( k \right) = \left[ {\begin{array}{*{20}c}
   {x\left( k \right)} & {u_1 \left( {k - 1} \right)} & {u_2 \left( {k - 1} \right)} &  \cdots  & {u_1 \left( 0 \right)} & {u_2 \left( 0 \right)}  \\
\end{array}} \right]^T.
\end{equation}

Then, we can rewrite (\ref{eq:7}) as
\begin{equation}\label{eq:9}
z\left( {k + 1} \right) = C_1 \left( k \right)z\left( k \right) + D_1 u_1 \left( k \right),
\end{equation}
where
\begin{equation}\label{eq:10}
\begin{split}
 &D_1  = \left( {\begin{array}{*{20}c}
   {\Gamma _{1,0} }  \\
   I  \\
   0  \\
    \vdots   \\
   0  \\
\end{array}} \right), \\
 &C_1 \left( k \right) = \left( {\begin{array}{*{20}c}
   {\Phi  + \Gamma _{2,0} A_2 \left( k \right)} & {\Gamma _{1,1}  + \bar B_{1,1}^2 \left( k \right)} & {\Gamma _{2,1}  + \bar B_{2,1}^2 \left( k \right)} & {\bar B_{1,2}^2 \left( k \right)} & {\bar B_{2,2}^2 \left( k \right)} &  \cdots  & {\bar B_{1,k}^2 \left( k \right)} & {\bar B_{2,k}^2 \left( k \right)}  \\
   0 & 0 & 0 & 0 & 0 &  \cdots  & 0 & 0  \\
   {A_2 \left( k \right)} & {B_{1,1}^2 \left( k \right)} & {B_{2,1}^2 \left( k \right)} & {B_{1,2}^2 \left( k \right)} & {B_{2,2}^2 \left( k \right)} &  \cdots  & {B_{1,k}^2 \left( k \right)} & {B_{2,k}^2 \left( k \right)}  \\
   0 & I & 0 & 0 & 0 &  \cdots  & 0 & 0  \\
   0 & 0 & I & 0 & 0 &  \cdots  & 0 & 0  \\
    \vdots  &  \vdots  &  \vdots  &  \ddots  &  \vdots  &  \cdots  &  \vdots  &  \vdots   \\
\end{array}} \right), \\
 &{\rm{ }}\bar B_{i,j}^2 \left( k \right) = \Gamma _{2,0} B_{i,j}^2 \left( k \right),\ \ {\rm{ }}i = 1,\ {\rm{ }}2;\ {\rm{ }}j = 1,\ {\rm{ }}2, \cdots ,\ k, \\
 \end{split}
\end{equation}
with 0 and $I$ denoting the zero matrix and the identity matrix, respectively, $u_i \left( k \right),\ i = 1,\ 2$
are set to be zero when $k < 0$.

Then, the optimization problem for controller 1 in (\ref{eq:5}) can be rewritten as
\begin{equation}\label{eq:11}
\begin{split}
 \mathop {\min }\limits_{u_1 \left( k \right),\ {\rm{ }}k = 0,\ 1, \cdots ,\ N - 1}\ \  &{\rm{  }}J_{1,N}  = z^T \left( N \right)Q_N^1 z\left( N \right) + \sum\limits_{k = 0}^{N - 1} {\left\{ {\left( {\begin{array}{*{20}c}
   {z\left( k \right)}  \\
   {u_1 \left( k \right)}  \\
\end{array}} \right)^T \left( {\begin{array}{*{20}c}
   {Q_{1,1}^1 } & 0  \\
   0 & {R_1 }  \\
\end{array}} \right)\left( {\begin{array}{*{20}c}
   {z\left( k \right)}  \\
   {u_1 \left( k \right)}  \\
\end{array}} \right)} \right\}},  \\
 {\rm{                    }}s.t.\ \ &{\rm{   }}z\left( {k + 1} \right) = C_1 \left( k \right)z\left( k \right) + D_1 u_1 \left( k \right), \\
 \end{split}
\end{equation}
where
\begin{equation}\label{eq:12}
Q_N^1  = \left( {\begin{array}{*{20}c}
   {Q_{1,N} } & 0 &  \cdots  & 0  \\
   0 & 0 &  \cdots  & 0  \\
    \vdots  &  \vdots  &  \ddots  &  \vdots   \\
   0 & 0 &  \cdots  & 0  \\
\end{array}} \right),\ {\rm{ }}Q_{1,1}^1  = \left( {\begin{array}{*{20}c}
   {Q_1 } & 0 &  \cdots  & 0  \\
   0 & 0 &  \cdots  & 0  \\
    \vdots  &  \vdots  &  \ddots  &  \vdots   \\
   0 & 0 &  \cdots  & 0  \\
\end{array}} \right).
\end{equation}

Define
\begin{equation}\label{eq:13}
V_L^1  = \mathop {\min }\limits_{u_1 \left( k \right),\ {\rm{ }}k = L,\  L+1, \cdots ,\ N - 1} \left\{ {z^T \left( N \right)Q_N^1 z\left( N \right) + \sum\limits_{k = L}^{N - 1} {\left\{ {\left( {\begin{array}{*{20}c}
   {z\left( k \right)}  \\
   {u_1 \left( k \right)}  \\
\end{array}} \right)^T \left( {\begin{array}{*{20}c}
   {Q_{1,1}^1 } & 0  \\
   0 & {R_1 }  \\
\end{array}} \right)\left( {\begin{array}{*{20}c}
   {z\left( k \right)}  \\
   {u_1 \left( k \right)}  \\
\end{array}} \right)} \right\}} } \right\}.
\end{equation}

We next derive the expressions for $V_L^1$ for different $L$.

\subsubsection{L = N} When $L = N$, we have
\begin{equation}\label{eq:14}
V_N^1  = z^T \left( N \right)S^1 \left( N \right)z\left( N \right),
\end{equation}
with
\[
S^1 \left( N \right) = Q_N^1.
\]

\subsubsection{$L = N-1$}
When $L = N-1$, from (\ref{eq:9}), (\ref{eq:13}) and (\ref{eq:14}), we get
\begin{equation}\label{eq:15}
\begin{split}
 V_{N - 1}^1  &= \mathop {\min }\limits_{u_1 \left( {N - 1} \right)} \left\{ {\left( {\begin{array}{*{20}c}
   {z\left( {N - 1} \right)}  \\
   {u_1 \left( {N - 1} \right)}  \\
\end{array}} \right)^T \left( {\begin{array}{*{20}c}
   {Q_{1,1}^1 } & 0  \\
   0 & {R_1 }  \\
\end{array}} \right)\left( {\begin{array}{*{20}c}
   {z\left( {N - 1} \right)}  \\
   {u_1 \left( {N - 1} \right)}  \\
\end{array}} \right) + z^T \left( N \right)S^1 \left( N \right)z\left( N \right)} \right\} \\
 &{\rm{      }} = \mathop {\min }\limits_{u_1 \left( {N - 1} \right)} \left( {\begin{array}{*{20}c}
   {z\left( {N - 1} \right)}  \\
   {u_1 \left( {N - 1} \right)}  \\
\end{array}} \right)^T \left( {\begin{array}{*{20}c}
   {P_{1,1}^1 \left( {N - 1} \right)} & {\left( {P_{1,2}^1 \left( {N - 1} \right)} \right)^T }  \\
   {P_{1,2}^1 \left( {N - 1} \right)} & {P_{2,2}^1 \left( {N - 1} \right)}  \\
\end{array}} \right)\left( {\begin{array}{*{20}c}
   {z\left( {N - 1} \right)}  \\
   {u_1 \left( {N - 1} \right)}  \\
\end{array}} \right), \\
 \end{split}
\end{equation}
where
\begin{equation}\label{eq:16}
\begin{split}
 P_{1,1}^1 \left( {N - 1} \right) &= C_1^T \left( {N - 1} \right)S^1 \left( N \right)C_1 \left( {N - 1} \right) + Q_{1,1}^1  ,\\
 P_{1,2}^1 \left( {N - 1} \right) &= D_1^T S^1 \left( N \right)C_1 \left( {N - 1} \right) ,\\
 P_{2,2}^1 \left( {N - 1} \right) &= D_1^T S^1 \left( N \right)D_1  + R_1  .\\
 \end{split}
\end{equation}

The optimal solution to (\ref{eq:15}) is given by \cite{IRef20}
\begin{equation}\label{eq:17}
u_1 \left( {N - 1} \right) =  - L_1 \left( {N - 1} \right)z\left( {N - 1} \right),
\end{equation}
where
\begin{equation}\label{eq:18}
L_1 \left( {N - 1} \right) = \left( {P_{2,2}^1 \left( {N - 1} \right)} \right)^{ - 1} P_{1,2}^1 \left( {N - 1} \right).
\end{equation}

Similarly, we get the optimal control strategy of the other controller as
\begin{equation}\label{eq:19}
u_2 \left( {N - 1} \right) =  - L_2 \left( {N - 1} \right)z\left( {N - 1} \right),
\end{equation}
where
\begin{equation}\label{eq:20}
L_2 \left( {N - 1} \right) = \left( {P_{2,2}^2 \left( {N - 1} \right)} \right)^{ - 1} P_{1,2}^2 \left( {N - 1} \right),
\end{equation}
and
\begin{equation}\label{eq:21}
\begin{split}
 &P_{1,2}^2 \left( {N - 1} \right) = D_2^T S^2 \left( N \right)C_2 \left( {N - 1} \right) ,\\
 &P_{2,2}^2 \left( {N - 1} \right) = D_2^T S^2 \left( N \right)D_2  + R_2  ,\\
 &S^2 \left( N \right) = \left( {\begin{array}{*{20}c}
   {Q_{2,N} } & 0 &  \cdots  & 0  \\
   0 & 0 &  \cdots  & 0  \\
    \vdots  &  \vdots  &  \ddots  &  \vdots   \\
   0 & 0 &  \cdots  & 0  \\
\end{array}} \right),\ {\rm{ }}D_2  = \left( {\begin{array}{*{20}c}
   {\Gamma _{2,0} }  \\
   0  \\
   I  \\
   0  \\
    \vdots   \\
   0  \\
\end{array}} \right) ,\\
 &C_2 \left( k \right) = \left( {\begin{array}{*{20}c}
   {\Phi  + \Gamma _{1,0} A_1 \left( k \right)} & {\Gamma _{1,1}  + \bar B_{1,1}^1 \left( k \right)} & {\Gamma _{2,1}  + \bar B_{2,1}^1 \left( k \right)} & {\bar B_{1,2}^1 \left( k \right)} & {\bar B_{2,2}^1 \left( k \right)} &  \cdots  & {\bar B_{1,k}^1 \left( k \right)} & {\bar B_{2,k}^1 \left( k \right)}  \\
   {A_1 \left( k \right)} & {B_{1,1}^1 \left( k \right)} & {B_{2,1}^1 \left( k \right)} & {B_{1,2}^1 \left( k \right)} & {B_{2,2}^1 \left( k \right)} &  \cdots  & {B_{1,k}^1 \left( k \right)} & {B_{2,k}^1 \left( k \right)}  \\
   0 & 0 & 0 & 0 & 0 &  \cdots  & 0 & 0  \\
   0 & I & 0 & 0 & 0 &  \cdots  & 0 & 0  \\
   0 & 0 & I & 0 & 0 &  \cdots  & 0 & 0  \\
    \vdots  &  \vdots  &  \vdots  &  \ddots  &  \vdots  &  \cdots  &  \vdots  &  \vdots   \\
\end{array}} \right) ,\\
 &{\rm{ }}\bar B_{i,j}^1 \left( k \right) = \Gamma _{1,0} B_{i,j}^1 \left( k \right),\ \ {\rm{ }}i = 1,\ {\rm{ }}2; \ {\rm{ }}j = 1,\ {\rm{ }}2, \cdots ,\ k{\rm{ }} .\\
 \end{split}
\end{equation}

From (\ref{eq:6}), (\ref{eq:8}), (\ref{eq:17}) and (\ref{eq:19}), we have
\begin{equation}\label{eq:22}
\begin{split}
 L_i \left( {N - 1} \right) =  &- \left[ {\begin{array}{*{20}c}
   {A_i \left( {N - 1} \right)} & {B_{1,1}^i \left( {N - 1} \right)} & {B_{2,1}^i \left( {N - 1} \right)}  \\
\end{array}} \right. \\
 {\rm{                       }}&\ \ \ \ \left. {\begin{array}{*{20}c}
    \cdots  & {B_{1,N - 1}^i \left( {N - 1} \right)} & {B_{2,N - 1}^i \left( {N - 1} \right)}  \\
\end{array}} \right],\ {\rm{ }}i = 1,\ {\rm{ }}2. \\
 \end{split}
\end{equation}

Based on (\ref{eq:18}), (\ref{eq:20}), from (\ref{eq:22}), we obtain
\begin{equation}\label{eq:23}
\begin{split}
 &B_{j,i}^1 \left( {N - 1} \right) = - \alpha B_{j,i}^2 \left( {N - 1} \right) ,\\
 &B_{j,i}^2 \left( {N - 1} \right) = - \beta B_{j,i}^1 \left( {N - 1} \right),\ {\rm{  }}i = 2,\ {\rm{ }}3, \cdots ,\ {\rm{ }}N - 1;\ {\rm{  }}j = 1,\ {\rm{ }}2 ,\\
 \end{split}
\end{equation}
where
\begin{equation}\label{eq:24}
\begin{split}
 \alpha & = \left( {\Gamma _{1,0}^T Q_{1,N} \Gamma _{1,0}  + R_1 } \right)^{ - 1} \Gamma _{1,0}^T Q_{1,N} \Gamma _{2,0}, \\
 \beta & = \left( {\Gamma _{2,0}^T Q_{2,N} \Gamma _{2,0}  + R_2 } \right)^{ - 1} \Gamma _{2,0}^T Q_{2,N}  \Gamma _{1,0} .\\
 \end{split}
\end{equation}

Then, from (\ref{eq:23}), we have
\begin{equation}\label{eq:25}
B_{j,i}^1 \left( {N - 1} \right) \equiv \alpha \beta B_{j,i}^1 \left( {N - 1} \right),\ {\rm{  }}i = 2,\ {\rm{ }}3, \cdots ,\ {\rm{ }}N - 1; \ {\rm{  }}j = 1,\ {\rm{ }}2,
\end{equation}
which means that
\begin{equation}\label{eq:26}
B_{j,i}^1 \left( {N - 1} \right) = B_{j,i}^2 \left( {N - 1} \right) = 0,\ \ {\rm{  }}i = 2,\ {\rm{ }}3, \cdots ,\ {\rm{ }}N - 1; \ {\rm{  }}j = 1,\ {\rm{ }}2.
\end{equation}

Based on (\ref{eq:22}) and (\ref{eq:26}), when $L = N - 1$, the optimal solutions can be simplified as
\begin{equation}\label{eq:27}
L_i \left( {N - 1} \right) =  - \left[ {\begin{array}{*{20}c}
   {A_i \left( {N - 1} \right)} & {B_{1,1}^i \left( {N - 1} \right)} & {B_{2,1}^i \left( {N - 1} \right)}  \\
\end{array}} \right],\ {\rm{ }}i = 1,\ {\rm{ }}2.
\end{equation}

Then, from (\ref{eq:17}) and (\ref{eq:19}), the optimal control strategies of the two controllers can be rewritten as
\begin{equation}\label{eq:28}
u_i \left( {N - 1} \right) =  - L_i \left( {N - 1} \right)\left( {\begin{array}{*{20}c}
   {x\left( {N - 1} \right)}  \\
   {u_1 \left( {N - 2} \right)}  \\
   {u_2 \left( {N - 2} \right)}  \\
\end{array}} \right),\ i = 1,\ 2,
\end{equation}
and the related parameters can be simplified as
\begin{equation}\label{eq:29}
Q_{1,1}^i  = \left( {\begin{array}{*{20}c}
   {Q_i } & 0 & 0  \\
   0 & 0 & 0  \\
   0 & 0 & 0  \\
\end{array}} \right),\ {\rm{ }}S^i \left( N \right) = \left( {\begin{array}{*{20}c}
   {Q_{i,N} } & 0 & 0  \\
   0 & 0 & 0  \\
   0 & 0 & 0  \\
\end{array}} \right),\ i = 1,\ 2,
\end{equation}
and
\begin{equation}\label{eq:30}
\begin{split}
 D_1  &= \left( {\begin{array}{*{20}c}
   {\Gamma _{1,0} }  \\
   I  \\
   0  \\
\end{array}} \right),\ {\rm{ }}D_2  = \left( {\begin{array}{*{20}c}
   {\Gamma _{2,0} }  \\
   0  \\
   I  \\
\end{array}} \right),{\rm{ }} \\
 C_1 \left( k \right) &= \left( {\begin{array}{*{20}c}
   {\Phi  + \Gamma _{2,0} A_2 \left( k \right)} & {\Gamma _{1,1}  + \Gamma _{2,0} B_{1,1}^2 \left( k \right)} & {\Gamma _{2,1}  + \Gamma _{2,0} B_{2,1}^2 \left( k \right)}  \\
   0 & 0 & 0  \\
   {A_2 \left( k \right)} & {B_{1,1}^2 \left( k \right)} & {B_{2,1}^2 \left( k \right)}  \\
\end{array}} \right), \\
 C_2 \left( k \right) &= \left( {\begin{array}{*{20}c}
   {\Phi  + \Gamma _{1,0} A_1 \left( k \right)} & {\Gamma _{1,1}  + \Gamma _{1,0} B_{1,1}^1 \left( k \right)} & {\Gamma _{2,1}  + \Gamma _{1,0} B_{2,1}^1 \left( k \right)}  \\
   {A_1 \left( k \right)} & {B_{1,1}^1 \left( k \right)} & {B_{2,1}^1 \left( k \right)}  \\
   0 & 0 & 0  \\
\end{array}} \right). \\
 \end{split}
\end{equation}

Substituting $u_1 \left( {N - 1} \right)$ in (\ref{eq:28}) into (\ref{eq:15}), $V_{N - 1}^1$ can be expressed as
\begin{equation}\label{eq:31}
V_{N - 1}^1  = \left( {\begin{array}{*{20}c}
   {x\left( {N - 1} \right)}  \\
   {u_1 \left( {N - 2} \right)}  \\
   {u_2 \left( {N - 2} \right)}  \\
\end{array}} \right)^T \bar S^1 \left( {N - 1} \right)\left( {\begin{array}{*{20}c}
   {x\left( {N - 1} \right)}  \\
   {u_1 \left( {N - 2} \right)}  \\
   {u_2 \left( {N - 2} \right)}  \\
\end{array}} \right) = z^T \left( {N - 1} \right)S^1 \left( {N - 1} \right)z\left( {N - 1} \right),
\end{equation}
where
\begin{equation}\label{eq:32}
\begin{split}
 &\bar S^1 \left( {N - 1} \right) = P_{1,1}^1 \left( {N - 1} \right) - L_1^T \left( {N - 1} \right)P_{2,2}^1 \left( {N - 1} \right)L_1 \left( {N - 1} \right) ,\\
 &S^1 \left( N \right) = \left( {\begin{array}{*{20}c}
   {\bar S^1 \left( {N - 1} \right)} & 0 &  \cdots  & 0  \\
   0 & 0 &  \cdots  & 0  \\
    \vdots  &  \vdots  &  \ddots  &  \vdots   \\
   0 & 0 &  \cdots  & 0  \\
\end{array}} \right) .\\
 \end{split}
\end{equation}

\subsubsection{$L = N - 2, \cdots ,\ 1,\ 0$}
When $L = N - 2$, from (\ref{eq:13}) and (\ref{eq:31}), we have
\begin{equation}\label{eq:33}
V_{N - 2}^1  = \mathop {\min }\limits_{u_1 \left( {N - 2} \right)} \left\{ {\left( {\begin{array}{*{20}c}
   {z\left( {N - 2} \right)}  \\
   {u_1 \left( {N - 2} \right)}  \\
\end{array}} \right)^T \left( {\begin{array}{*{20}c}
   {Q_{1,1}^1 } & 0  \\
   0 & {R_1 }  \\
\end{array}} \right)\left( {\begin{array}{*{20}c}
   {z\left( {N - 2} \right)}  \\
   {u_1 \left( {N - 2} \right)}  \\
\end{array}} \right) + z^T \left( {N - 1} \right)S^1 \left( {N - 1} \right)z\left( {N - 1} \right)} \right\}.
\end{equation}

We can see that, (\ref{eq:15}) and (\ref{eq:33}) have the same form.
Thus, repeat the same process as that for $L = N - 1$, we can derive the optimal
controller $u_1 \left( k \right),\ {\rm{  }}k = N - 2, \cdots ,\ {\rm{ 1}},\ {\rm{ }}0{\rm{ }}$,
which can be expressed as
\begin{equation}\label{eq:34}
u_i \left( k \right) =  - L_i \left( k \right)\left( {\begin{array}{*{20}c}
   {x\left( k \right)}  \\
   {u_1 \left( {k - 1} \right)}  \\
   {u_2 \left( {k - 1} \right)}  \\
\end{array}} \right),\ \ {\rm{  }}i = 1,\ {\rm{ }}2;\ {\rm{ }}k = 0,\ {\rm{ }}1,{\rm{ }} \cdots ,\ {\rm{ }}N - 1,
\end{equation}
where
\begin{equation}\label{eq:35}
\begin{split}
 L_i \left( k \right) &= \left( {P_{2,2}^i \left( k \right)} \right)^{ - 1} P_{1,2}^i \left( k \right) ,\\
 S^i \left( k \right) &= P_{1,1}^i \left( k \right) - L_i^T \left( k \right)P_{2,2}^i \left( k \right)L_i \left( k \right){\rm{ }} ,\\
 \end{split}
\end{equation}
and
\begin{equation}\label{eq:36}
\begin{split}
 P_{1,1}^i \left( k \right) &= C_i^T \left( k \right)S^i \left( {k + 1} \right)C_i \left( k \right) + Q_{1,1}^i  ,\\
 P_{1,2}^i \left( k \right) &= D_i^T S^i \left( {k + 1} \right)C_i \left( k \right) ,\\
 P_{2,2}^i \left( k \right) &= D_i^T S^i \left( {k + 1} \right)D_i  + R_i  .\\
 \end{split}
\end{equation}

From (\ref{eq:6}) and (\ref{eq:34}), we have
\begin{equation}\label{eq:37}
L_i \left( k \right) =  - \left[ {\begin{array}{*{20}c}
   {A_i \left( k \right)} & {B_{1,1}^i \left( k \right)} & {B_{2,1}^i \left( k \right)}  \\
\end{array}} \right],\ i = 1,\ 2;\ k = 0,\ {\rm{ }}1,{\rm{ }} \cdots ,\ {\rm{ }}N - 1,
\end{equation}
which means that the optimal control strategies are the linear with current plant states and
the last control strategies.

Based on (\ref{eq:35}) and (\ref{eq:37}), we can deduce the values
of $A_i \left( k \right)$, $B_{1,1}^i \left( k \right)$ and $B_{2,1}^i \left( k \right)$, $i = 1,\ 2$
as follows (see Appendix A for details).
\begin{equation}\label{eq:38}
\begin{split}
 A_1 \left( k \right) &= \left[ {I - a_2^1 \left( k \right)a_2^2 \left( k \right)} \right]^{ - 1} \left[ {a_2^1 \left( k \right)a_1^2 \left( k \right) - a_1^1 \left( k \right)} \right], \\
 B_{1,1}^1 \left( k \right) &= \left[ {I - b_2^1 \left( k \right)b_2^2 \left( k \right)} \right]^{ - 1} \left[ {b_2^1 \left( k \right)b_1^2 \left( k \right) - b_1^1 \left( k \right)} \right], \\
 B_{2,1}^1 \left( k \right) &= \left[ {I - c_2^1 \left( k \right)c_2^2 \left( k \right)} \right]^{ - 1} \left[ {c_2^1 \left( k \right)c_1^2 \left( k \right) - c_1^1 \left( k \right)} \right], \\
 A_2 \left( k \right) &= \left[ {I - a_2^2 \left( k \right)a_2^1 \left( k \right)} \right]^{ - 1} \left[ {a_2^2 \left( k \right)a_1^1 \left( k \right) - a_1^2 \left( k \right)} \right], \\
 B_{1,1}^2 \left( k \right) &= \left[ {I - b_2^2 \left( k \right)b_2^1 \left( k \right)} \right]^{ - 1} \left[ {b_2^2 \left( k \right)b_1^1 \left( k \right) - b_1^2 \left( k \right)} \right], \\
 B_{2,1}^2 \left( k \right) &= \left[ {I - c_2^2 \left( k \right)c_2^1 \left( k \right)} \right]^{ - 1} \left[ {c_2^2 \left( k \right)c_1^1 \left( k \right) - c_1^2 \left( k \right)} \right]. \\
 \end{split}
\end{equation}

Then, using (\ref{eq:37}) and (\ref{eq:38}), from (\ref{eq:34}), we can achieve the optimal control strategies.
The algorithm can be summarized as follows.

\begin{minipage}[h]{6.2 in}
\rule{\linewidth}{0.5mm}\vspace{-.1in}
{\bf {\footnotesize The optimal distributed controllers}}\vspace{-.1in}\\
\rule{\linewidth}{0.3mm}
{ {
\begin{tabular}{ll}
    \;\textbf {Off-line}: \\
    \;1: Initialize $S^1 \left( N \right)$ and $S^2 \left( N \right)$ using (\ref{eq:29}).\\
    \;2:  \textbf {for $k = N - 1: - 1:0$ do}.\\
    \;3:\ \ Calculate $A_i \left( k \right)$, $B_{1,1}^i \left( k \right)$ and $B_{2,1}^i \left( k \right)$, $i = 1,\ 2$ using (\ref{eq:38}).\\
    \;\ \ \ \ Calculate $L_1 \left( k \right)$ and $L_2 \left( k \right)$ using (\ref{eq:37}).\\
    \;\ \ \ \ Calculate $S^1 \left( k \right)$ and $S^2 \left( k \right)$ using (\ref{eq:35}).\\
    \;4: \textbf{end for}.\\
    \;\textbf {On-line}: \\
    \;1: Initialize $x\left( 0 \right) = x_0$, and $u_i \left( k \right) = 0,\ {\rm{ }}i = 1,\ {\rm{ }}2;\ {\rm{ }}k < 0$. \\
    \;2:  \textbf {for $k = 0: 1:N - 1$ do} .\\
    \;3:\ \ Use $x\left( k \right)$, $u_1 \left( {k - 1} \right)$, $u_2 \left( {k - 1} \right)$ and $L_1 \left( k \right)$ to compute $u_1 \left( k \right)$ in (\ref{eq:34}) .\\
    \;\ \ \ \ Use $x\left( k \right)$, $u_1 \left( {k - 1} \right)$, $u_2 \left( {k - 1} \right)$ and $L_2 \left( k \right)$ to compute $u_2 \left( k \right)$ in (\ref{eq:34})  .\\
    \;\ \ \ \ Exchange control signals $u_1 \left( k \right)$ and $u_2 \left( k \right)$ between the two controllers.\\
    \;4: \textbf{end for}.\\
\end{tabular}}}\\
\rule{\linewidth}{0.3mm}
\end{minipage}\vspace{.2 in}\\

\subsection{Special Cases}

In this subsection, the optimal control solutions derived in the last subsection are applied to
two special cases. One is the optimal control strategy for a single controller with time delay,
and the other is the optimal control strategy for two controllers without time delays.

\subsubsection{Single controller with time delay}\

Consider the case of a single controller, we have
\begin{equation}\label{eq:39}
\begin{split}
 &A_2 \left( k \right) = B_{1,1}^2 \left( k \right) = B_{2,1}^2 \left( k \right) = 0 ,\\
 &\Gamma _{2,0}  = \Gamma _{2,1}  = 0 .\\
 \end{split}
\end{equation}

Then, the optimal control strategies in (\ref{eq:34}) can be simplified as
\begin{equation}\label{eq:40}
u_1 \left( k \right) =  - L_1 \left( k \right)\left( {\begin{array}{*{20}c}
   {x\left( k \right)}  \\
   {u_1 \left( {k - 1} \right)}  \\
\end{array}} \right),\ \ {\rm{  }}k = 0,\ {\rm{ }}1,{\rm{ }} \cdots ,\ {\rm{ }}N - 1,
\end{equation}
where
\begin{equation}\label{eq:41}
\begin{split}
 L_1 \left( k \right) &= \left( {P_{2,2}^1 \left( k \right)} \right)^{ - 1} P_{1,2}^1 \left( k \right) ,\\
 S^1 \left( k \right) &= P_{1,1}^1 \left( k \right) - L_1^T \left( k \right)P_{2,2}^1 \left( k \right)L_1 \left( k \right) ,\\
 S^1 \left( N \right) &= \left( {\begin{array}{*{20}c}
   {Q_{1,N} } & 0  \\
   0 & 0  \\
\end{array}} \right) ,\\
 P_{1,1}^1 \left( k \right) &= C_1^T \left( k \right)S^1 \left( k \right)C_1 \left( k \right) + Q_{1,1}^1  ,\\
 P_{1,2}^1 \left( k \right) &= D_1^T S^1 \left( {k + 1} \right)C_1 \left( k \right) ,\\
 P_{2,2}^1 \left( k \right) &= D_1^T S^1 \left( {k + 1} \right)D_1  + R_1  ,\\
 \end{split}
\end{equation}
and
\begin{equation}\label{eq:42}
Q_{1,1}^1  = \left( {\begin{array}{*{20}c}
   {Q_1 } & 0  \\
   0 & 0  \\
\end{array}} \right),\ {\rm{ }}D_1  = \left( {\begin{array}{*{20}c}
   {\Gamma _{1,0} }  \\
   I  \\
\end{array}} \right),\ {\rm{ }}C_1 \left( k \right) = \left( {\begin{array}{*{20}c}
   \Phi  & {\Gamma _{1,1} }  \\
   0 & 0  \\
\end{array}} \right).
\end{equation}

The above optimal solution is the same as that in \cite{IRef19} when the time delay is deterministic.

\subsubsection{Two controllers without time delays}\

If we ignore the time delays, we have
\begin{equation}\label{eq:43}
\begin{split}
 &B_{1,1}^1 \left( k \right) = B_{2,1}^1 \left( k \right) = B_{1,1}^2 \left( k \right) = B_{2,1}^2 \left( k \right) = 0 ,\\
 &\Gamma _{i,1}  = 0,\ {\rm{  }}i = 1,\ {\rm{ }}2 .\\
 \end{split}
\end{equation}

Then, the optimal control strategies become
\begin{equation}\label{eq:44}
u_i \left( k \right) = A_i \left( k \right)x\left( k \right),\ {\rm{  }}i = 1,\ {\rm{ }}2;\ {\rm{ }}k = 0,\ {\rm{ }}1, \cdots ,\ {\rm{ }}N - 1,
\end{equation}
where $A_i \left( k \right)$ is derived by
\begin{equation}\label{eq:45}
\begin{split}
 A_1 \left( k \right) &= \left[ {I - a_2^1 \left( k \right)a_2^2 \left( k \right)} \right]^{ - 1} \left[ {a_2^1 \left( k \right)a_1^2 \left( k \right) - a_1^1 \left( k \right)} \right], \\
 A_2 \left( k \right) &= \left[ {I - a_2^2 \left( k \right)a_2^1 \left( k \right)} \right]^{ - 1} \left[ {a_2^2 \left( k \right)a_1^1 \left( k \right) - a_1^2 \left( k \right)} \right], \\
 \end{split}
\end{equation}
where
\begin{equation}\label{eq:56}
\begin{split}
 a_1^i \left( k \right) &= \left( {R_i  + \Gamma _{i,0}^T S^i \left( {k + 1} \right)\Gamma _{i,0} } \right)^{ - 1} \Gamma _{i,0}^T S^i \left( {k + 1} \right)\Phi , \\
 a_2^i \left( k \right) &= \left( {R_i  + \Gamma _{i,0}^T S^i \left( {k + 1} \right)\Gamma _{i,0} } \right)^{ - 1} \Gamma _{i,0}^T S^i \left( {k + 1} \right)\Gamma _{3 - i,0} , \\
 \end{split}
\end{equation}
and
\begin{equation}\label{eq:47}
\begin{split}
 S^i \left( N \right) &= Q_{i,N} , \\
 S^i \left( k \right) &= Q_i  + \left( {\Phi  + \Gamma _{3 - i,0} A_{3 - i} \left( k \right)} \right)^T S^i \left( {k + 1} \right)\left( {\Phi  + \Gamma _{3 - i,0} A_{3 - i} \left( k \right)} \right) \\
 &\ \ \ \ {\rm{            }} - A_i^T \left( k \right)\left( {\Gamma _{i,0}^T S^i \left( {k + 1} \right)\Gamma _{i,0}  + R_i } \right)A_i \left( k \right). \\
  \end{split}
\end{equation}

These results correspond to the discrete-time control strategies for the non-cooperative feedback games in \cite{IRef18}.

\section{Extension to Multiple Distributed Controllers}

In this section, we extend the results for the case of two distributed controllers in Section III
to multiple distributed controllers. We will omit the detailed derivations since they are similar
to those in Section III.

Similar to (\ref{eq:37}), the optimal linear control strategies are linear with the current
plant states and the last control strategies, i.e.,
\begin{equation}\label{eq:48}
u_i \left( k \right) = A_i \left( k \right)x\left( k \right) + \sum\limits_{j = 1}^p {B_j^i \left( k \right)u_j \left( {k - 1} \right)} ,\ {\rm{ }}i = 1,\ {\rm{ }}2,{\rm{ }} \cdots ,\ {\rm{ }}p,
\end{equation}
where $p$ is the number of controllers, $A_i$ and $B_j^i$, $j = 1,\ 2, \cdots ,\ p$ are coefficient matrices.

Taking controller $i$ as the desired one, we can rewrite the control process as
\begin{equation}\label{eq:49}
\begin{split}
 x\left( {k + 1} \right) &= \Phi x\left( k \right) + \sum\limits_{j = 1}^p {\left[ {\Gamma _{j,0} u_j \left( k \right) + \Gamma _{j,1} u_j \left( {k - 1} \right)} \right]}  \\
 {\rm{           }} &= \left( {\Phi  + \sum\limits_{\scriptstyle m = 1 \hfill \atop
  \scriptstyle m \ne i \hfill}^p {\Gamma _{m,0} A_m \left( k \right)} } \right)x\left( k \right) + \Gamma _{i,0} u_i \left( k \right) + \sum\limits_{j = 1}^p {\left[ {\left( {\Gamma _{j,1}  + \sum\limits_{\scriptstyle n = 1 \hfill \atop
  \scriptstyle n \ne i \hfill}^p {\Gamma _{n,0} B_j^n \left( k \right)} } \right)u_j \left( {k - 1} \right)} \right]} . \\
 \end{split}
\end{equation}

Define
\begin{equation}\label{eq:50}
z\left( k \right) = \left( {\begin{array}{*{20}c}
   {x\left( k \right)}  \\
   {u_1 \left( {k - 1} \right)}  \\
   {u_2 \left( {k - 1} \right)}  \\
    \vdots   \\
   {u_p \left( {k - 1} \right)}  \\
\end{array}} \right).
\end{equation}

Similarly as in Section III, we can derive the optimal solution for controller $i$ as
\begin{equation}\label{eq:51}
\begin{split}
 u_i \left( k \right) &=  - L_i \left( k \right)z\left( k \right),\ {\rm{  }}k = 0,\ {\rm{ }}1,{\rm{ }} \cdots ,\ {\rm{ }}N - 1, \\
 L_i \left( k \right) &= \left( {P_{2,2}^i \left( k \right)} \right)^{ - 1} P_{1,2}^i \left( k \right), \\
 \end{split}
\end{equation}
where
\begin{equation}\label{eq:52}
\begin{split}
 S^i \left( k \right) &= P_{1,1}^i \left( k \right) - L_i^T \left( k \right)P_{2,2}^i \left( k \right)L_i \left( k \right) ,\\
 S^i \left( N \right) &= \left( {\begin{array}{*{20}c}
   {Q_{i,N} } & 0 &  \cdots  & 0  \\
   0 & 0 &  \cdots  & 0  \\
    \vdots  &  \vdots  &  \ddots  &  \vdots   \\
   0 & 0 &  \cdots  & 0  \\
\end{array}} \right) ,\\
 P_{1,1}^i \left( k \right) &= C_i^T \left( k \right)S^i \left( {k + 1} \right)C_i \left( k \right) + Q_{1,1}^i  ,\\
 P_{1,2}^i \left( k \right) &= D_i^T S^i \left( {k + 1} \right)C_i \left( k \right) ,\\
 P_{2,2}^i \left( k \right) &= D_i^T S^i \left( {k + 1} \right)D_i  + R_i  ,\\
 \end{split}
\end{equation}
and
\[
Q_{1,1}^i  = \left( {\begin{array}{*{20}c}
   {Q_i } & 0 &  \cdots  & 0  \\
   0 & 0 &  \cdots  & 0  \\
    \vdots  &  \vdots  &  \ddots  & 0  \\
   0 & 0 & 0 & 0  \\
\end{array}} \right),\ {\rm{ }}D_i  = \left( {\begin{array}{*{20}c}
   {\Gamma _{i,0} }  \\
   0  \\
    \vdots   \\
   0  \\
   {I_{i + 1} }  \\
   0  \\
    \vdots   \\
   0  \\
\end{array}} \right),
\]
\begin{equation}\label{eq:53}
C_i \left( k \right) = \left( {\begin{array}{*{20}c}
   {\left( {\Phi  + \sum\limits_{\scriptstyle n = 1 \hfill \atop
  \scriptstyle n \ne i \hfill}^p {\Gamma _{n,0} A_n \left( k \right)} } \right)} & {\left( {\Gamma _{1,1}  + \bar B_1 \left( k \right)} \right)} & {\left( {\Gamma _{2,1}  + \bar B_2 \left( k \right)} \right)} &  \cdots  & {\left( {\Gamma _{p,1}  + \bar B_p \left( k \right)} \right)}  \\
   {A_1 \left( k \right)} & {B_1^1 \left( k \right)} & {B_2^1 \left( k \right)} &  \cdots  & {B_p^1 \left( k \right)}  \\
    \vdots  &  \vdots  &  \vdots  &  \ddots  &  \vdots   \\
   {A_{i - 1} \left( k \right)} & {B_1^{i - 1} \left( k \right)} & {B_2^{i - 1} \left( k \right)} &  \cdots  & {B_p^{i - 1} \left( k \right)}  \\
   0 & 0 & 0 & 0 & 0  \\
   {A_{i + 1} \left( k \right)} & {B_1^{i + 1} \left( k \right)} & {B_2^{i + 1} \left( k \right)} &  \cdots  & {B_p^{i + 1} \left( k \right)}  \\
    \vdots  &  \vdots  &  \vdots  &  \ddots  &  \vdots   \\
   {A_p \left( k \right)} & {B_1^p \left( k \right)} & {B_2^p \left( k \right)} &  \cdots  & {B_p^p \left( k \right)}  \\
\end{array}} \right),
\end{equation}
that $\bar B_j \left( k \right) = \sum\limits_{\scriptstyle n = 1 \hfill \atop
  \scriptstyle n \ne i \hfill}^p {\Gamma _{n,0} B_j^n \left( k \right)}$,
and $I_{i + 1}$ denotes the $(i + 1)$-$th$ block of $D_i$, which is a identity matrix of size $N \times N$.

From (\ref{eq:48}) and (\ref{eq:51}), for controller $i$, ${\rm{ }}i = 1,\ {\rm{ 2}}, \cdots ,\ p$, we can obtain
\begin{equation}\label{eq:54}
\begin{split}
 A_i \left( k \right) &= E_i^{ - 1} \left[ {\Gamma _{i,0}^T S_{1,1}^i \left( {k + 1} \right)\Phi  + S_{i + 1,1}^i \left( {k + 1} \right)\Phi  + \sum\limits_{\scriptstyle j = 1 \hfill \atop
  \scriptstyle j \ne i \hfill}^p {F_i^j A_j \left( k \right)} } \right], \\
 B_1^i \left( k \right) &= E_i^{ - 1} \left[ {\Gamma _{i,0}^T S_{1,1}^i \left( {k + 1} \right)\Gamma _{1,1}  + S_{i + 1,1}^i \left( {k + 1} \right)\Gamma _{1,1}  + \sum\limits_{\scriptstyle j = 1 \hfill \atop
  \scriptstyle j \ne i \hfill}^p {F_i^j B_1^j \left( k \right)} } \right], \\
 {\rm{          }} &\ \ \quad \quad \quad \quad \quad \quad \quad \quad \quad \vdots  \\
 B_p^i \left( k \right) &= E_i^{ - 1} \left[ {\Gamma _{i,0}^T S_{1,1}^i \left( {k + 1} \right)\Gamma _{p,1}  + S_{i + 1,1}^i \left( {k + 1} \right)\Gamma _{p,1}  + \sum\limits_{\scriptstyle j = 1 \hfill \atop
  \scriptstyle j \ne i \hfill}^p {F_i^j B_p^j \left( k \right)} } \right], \\
 \end{split}
\end{equation}
where
\begin{equation}\label{eq:55}
\begin{split}
 E_i  &= D_i^T S^i \left( {k + 1} \right)D_i  + R_i , \\
 F_i^j  &= \Gamma _{i,0}^T S_{1,1}^i \left( {k + 1} \right)\Gamma _{j,0}  + S_{i + 1,1}^i \left( {k + 1} \right)\Gamma _{j,0}  + \Gamma _{i,0}^T S_{1,j + 1}^i \left( {k + 1} \right) + S_{i + 1,j + 1}^i \left( {k + 1} \right), \\
 \end{split}
\end{equation}
and $S_{m,n}^i \left( {k + 1} \right)$ is the $\left( {m,n} \right)$-$th$ block of matrix $S^i \left( {k + 1} \right)$,
whose size is $M \times M$ when $m = n = 1$, $M \times N$ when $m = 1$; $n \ge 2$, $N \times M$ when $m \ge 2$; $n = 1$,
and $N \times N$ when $m \ge 2$; $n \ge 2$.

It can be seen that all the equations in (\ref{eq:54}) are linear functions. We can easily calculate all values
of $A_i \left( k \right),\ {\rm{ }}B_j^i \left( k \right),\ {\rm{ }}i = 1,\ {\rm{ }}2, \cdots ,\ p,\ {\rm{ }}j = 1,\ {\rm{ }}2, \cdots ,\ p$.
Then we can obtain the optimal control strategies from (\ref{eq:48}).

\begin{figure}[h]
  \begin{center}
  \includegraphics[width=0.8\textwidth]{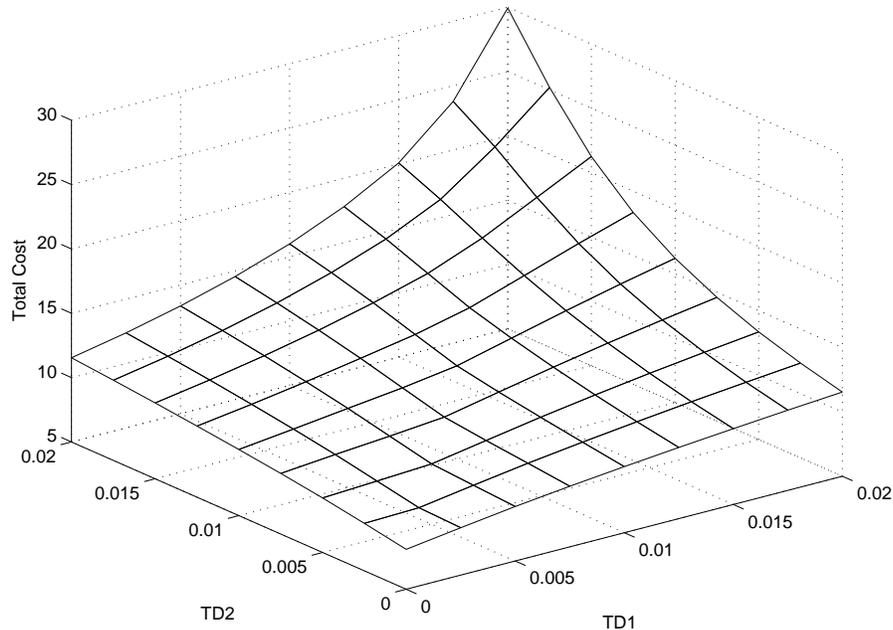}
  \caption{Total cost with two distributed controllers under various time delays.} \label{Fig3}
  \end{center}
\end{figure}

\section{Simulation Results}

In this section, we provide simulation studies on the performance of the proposed optimal
distributed control strategies.
First we consider a generic control system, and then introduce a power-grid application.

\subsection{A Generic System}

We consider a system with two distributed controllers.
The sampling period is chosen as $h=0.05$, the sampling duration $N=50$,
and the other parameters of the control system are set as follows \cite{IRef7}:
\[
A = \left[ {\begin{array}{*{20}c}
   0 & 1  \\
   { - 3} & { - 4}  \\
\end{array}} \right],{\rm{ }}\ \ B_1  = B_2  = \left[ {\begin{array}{*{20}c}
   0  \\
   1  \\
\end{array}} \right],
\]
and, for simplicity, we choose
\[
Q_{i,N}  = Q_i  = \left[ {\begin{array}{*{20}c}
   1 & 0  \\
   0 & 1  \\
\end{array}} \right] \times 100,\ {\rm{ }}R_i  = 1,\ {\rm{  }}i = 1,\ 2.
\]

Fig.~\ref{Fig3} shows the total costs of the system with various time delays.
TD1 and TD2 represent the time delays of controller 1 and controller 2, respectively.
It can be seen that the total cost becomes larger when either time delay increases,
especially when both time delays are large. This is because
the effect of the previous control signals increases when the time delay becomes larger,
which leads to less system stability and larger cost.

\begin{figure}[h]
  \begin{center}
  \includegraphics[width=0.8\textwidth]{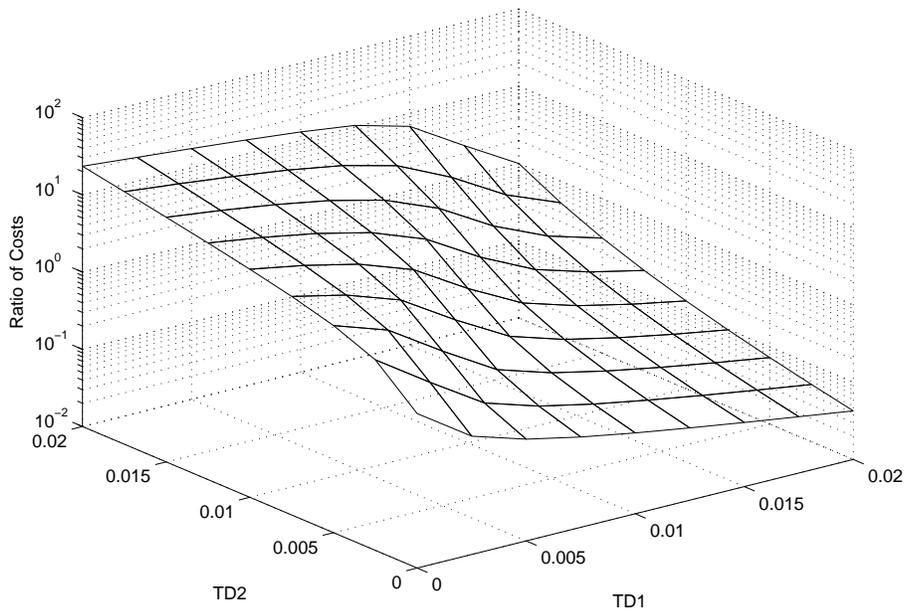}
  \caption{Ratio of costs between two distributed controllers under various time delays.} \label{Fig4}
  \end{center}
\end{figure}

Fig.~\ref{Fig4} depicts the ratio of costs between controller 1 and controller 2.
It can be seen that the ratio decreases with TD1 and increases with TD2,
which means that the controller with smaller time delay contributes more to the total cost.
This is because when the time delay becomes smaller,
the system becomes more stable using the corresponding controller,
and the effect of the controller will increase.

\begin{figure}[h]
  \begin{center}
  \includegraphics[width=0.9\textwidth]{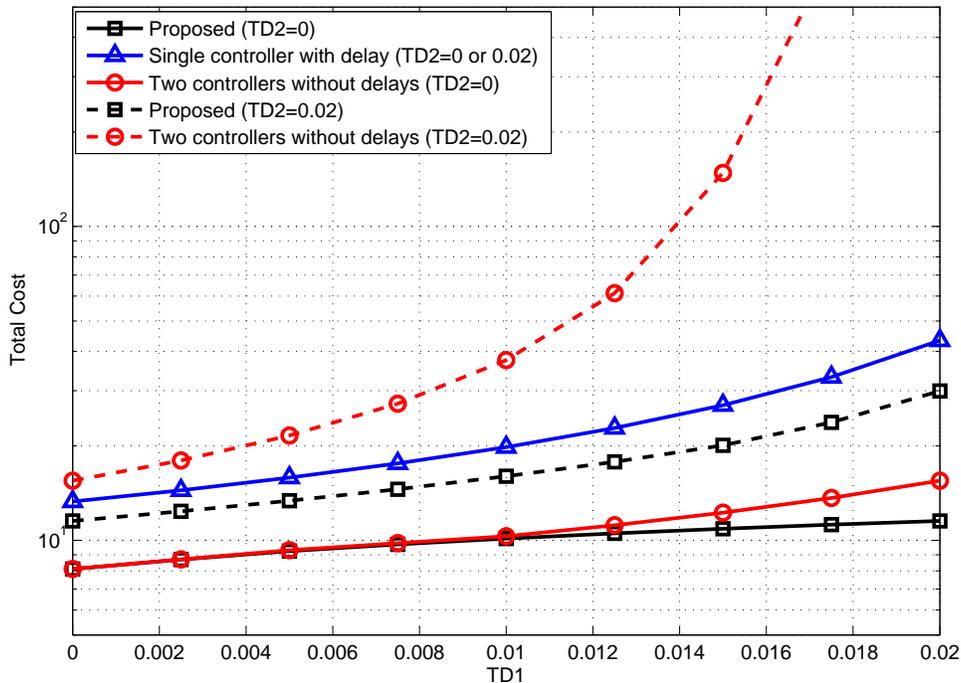}
  \caption{Performance comparison for three schemes with various TD1.} \label{Fig5}
  \end{center}
\end{figure}

Fig.~\ref{Fig5} shows the performance comparison for three schemes:
(1) the proposed distributed control algorithm;
(2) the LQG-controller algorithm designed for a single controller with time delay in \cite{IRef19}, where
controller 1 is assumed to be the desired controller;
(3) the LQG-controller algorithm designed for two distributed controllers neglecting the time delays
in \cite{IRef18}.
In the simulations, TD2 is set to be 0 and 0.02, and TD1 varies within [0, 0.02].
It can be seen that the proposed algorithm outperforms the other two schemes
in the sense that it has a lower total cost.
It can also be seen that the total cost of the two distributed controllers without delays
increases more rapidly with the time delay than the other two schemes when TD2 = 0.02.
It is because this scheme cannot effectively deal with the time delay so that large time delay
introduces severe performance degradation.
Note that, in Fig.~\ref{Fig5}, the total cost of the single controller scheme
is the same for different TD2, since only controller 1 is considered in this scheme.

\subsection{Load Frequency Control in Power Grid}

\begin{figure}[h]
  \begin{center}
  \includegraphics[width=0.9\textwidth]{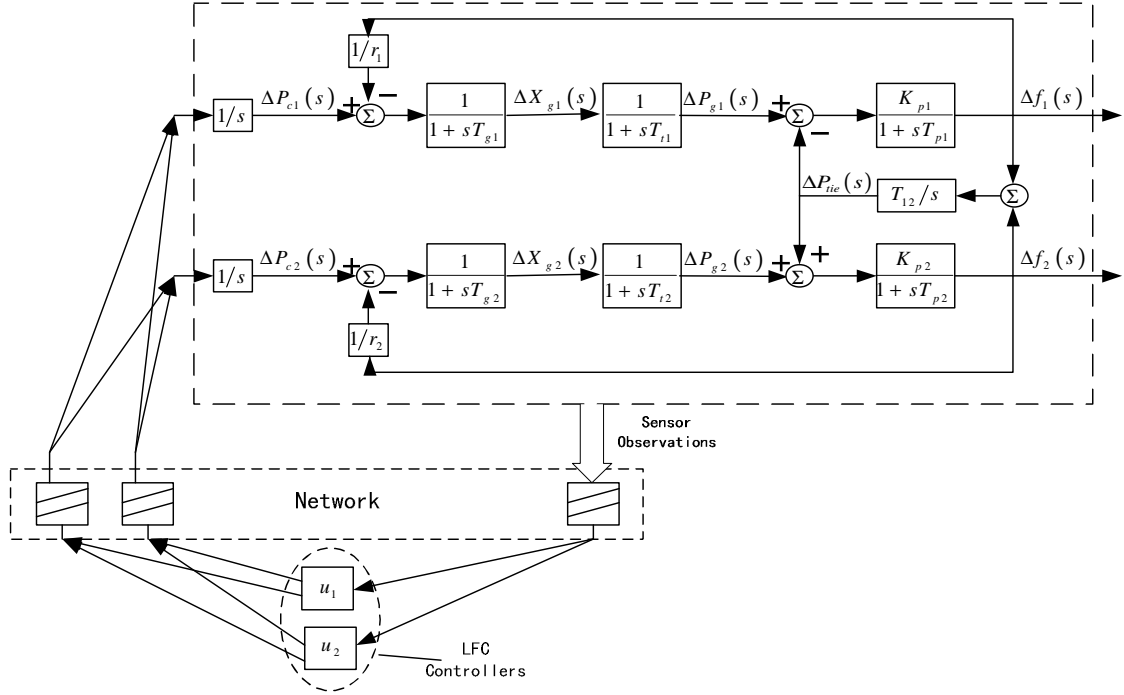}
  \caption{Block diagram of a two-area LFC system for power grid.} \label{Fig6}
  \end{center}
\end{figure}

We next consider the application of the proposed optimal distributed control scheme to
two-area load frequency control (LFC) in power grid systems \cite{IRef21}\cite{IRef22}.
The LFC block diagram is shown in Fig.~\ref{Fig6},
where the system states and feedback control signals are separately
transmitted through a shared wireless network, which incur the time delays.
The adjusting speed $u_i$ will be optimally designed according to the requested deviation of
generator outputs $\Delta P_{ci}$.

The linear dynamic control model can be described as
\begin{equation}\label{eq:56}
\dot x\left( t \right) = A^c x\left( t \right) + B_1^c u_1 \left( {t - \tau _1 } \right) + B_2^c u_2 \left( {t - \tau _2 } \right),
\end{equation}
where
\begin{equation}\label{eq:57}
\begin{split}
 &x\left( t \right) = \left[ {\begin{array}{*{20}c}
   {\Delta f_1 } & {\Delta P_{g1} } & {\Delta X_{g1} } & {\Delta f_2 } & {\Delta P_{g2} } & {\Delta X_{g2} } & {\Delta P_{tie} } & {\Delta P_{c1} } & {\Delta P_{c2} }  \\
\end{array}} \right]^T  ,\\
 &A^c = \left[ {\begin{array}{*{20}c}
   {{{ - 1} \mathord{\left/
 {\vphantom {{ - 1} {T_{p1} }}} \right.
 \kern-\nulldelimiterspace} {T_{p1} }}} & {{{K_{p1} } \mathord{\left/
 {\vphantom {{K_{p1} } {T_{p1} }}} \right.
 \kern-\nulldelimiterspace} {T_{p1} }}} & 0 & 0 & 0 & 0 & {{{K_{p1} } \mathord{\left/
 {\vphantom {{K_{p1} } {T_{p1} }}} \right.
 \kern-\nulldelimiterspace} {T_{p1} }}} & 0 & 0  \\
   0 & {{{ - 1} \mathord{\left/
 {\vphantom {{ - 1} {T_{t1} }}} \right.
 \kern-\nulldelimiterspace} {T_{t1} }}} & {{1 \mathord{\left/
 {\vphantom {1 {T_{t1} }}} \right.
 \kern-\nulldelimiterspace} {T_{t1} }}} & 0 & 0 & 0 & 0 & 0 & 0  \\
   {{{ - 1} \mathord{\left/
 {\vphantom {{ - 1} {r_1 T_{g1} }}} \right.
 \kern-\nulldelimiterspace} {r_1 T_{g1} }}} & 0 & {{{ - 1} \mathord{\left/
 {\vphantom {{ - 1} {T_{g1} }}} \right.
 \kern-\nulldelimiterspace} {T_{g1} }}} & 0 & 0 & 0 & 0 & {{1 \mathord{\left/
 {\vphantom {1 {T_{g1} }}} \right.
 \kern-\nulldelimiterspace} {T_{g1} }}} & 0  \\
   0 & 0 & 0 & {{{ - 1} \mathord{\left/
 {\vphantom {{ - 1} {T_{p2} }}} \right.
 \kern-\nulldelimiterspace} {T_{p2} }}} & {{{K_{p2} } \mathord{\left/
 {\vphantom {{K_{p2} } {T_{p2} }}} \right.
 \kern-\nulldelimiterspace} {T_{p2} }}} & 0 & {{{K_{p2} } \mathord{\left/
 {\vphantom {{K_{p2} } {T_{p2} }}} \right.
 \kern-\nulldelimiterspace} {T_{p2} }}} & 0 & 0  \\
   0 & 0 & 0 & 0 & {{{ - 1} \mathord{\left/
 {\vphantom {{ - 1} {T_{t2} }}} \right.
 \kern-\nulldelimiterspace} {T_{t2} }}} & {{1 \mathord{\left/
 {\vphantom {1 {T_{t2} }}} \right.
 \kern-\nulldelimiterspace} {T_{t2} }}} & 0 & 0 & 0  \\
   0 & 0 & 0 & {{{ - 1} \mathord{\left/
 {\vphantom {{ - 1} {r_2 T_{g2} }}} \right.
 \kern-\nulldelimiterspace} {r_2 T_{g2} }}} & 0 & {{{ - 1} \mathord{\left/
 {\vphantom {{ - 1} {T_{g2} }}} \right.
 \kern-\nulldelimiterspace} {T_{g2} }}} & 0 & 0 & {{1 \mathord{\left/
 {\vphantom {1 {T_{g2} }}} \right.
 \kern-\nulldelimiterspace} {T_{g2} }}}  \\
   {T_{12} } & 0 & 0 & { - T_{12} } & 0 & 0 & 0 & 0 & 0  \\
   0 & 0 & 0 & 0 & 0 & 0 & 0 & 0 & 0  \\
   0 & 0 & 0 & 0 & 0 & 0 & 0 & 0 & 0  \\
\end{array}} \right] ,\\
 &B_1^c  = B_2^c  = \left[ {\begin{array}{*{20}c}
   0 & 0 & 0 & 0 & 0 & 0 & 0 & 1 & 1  \\
\end{array}} \right] ,\\
 \end{split}
\end{equation}
and the subscript $i=1,\ 2$ representing the $i$-$th$ control area, $\Delta f_i$ is the deviation of frequency,
$\Delta P_{gi}$ is the deviation of generator mechanical output,
$\Delta X_{gi}$ is the deviation of valve position, $\Delta P_{tie}$ is the deviation of tie-line power,
$\Delta P_{ci}$ is the requested deviation of generator output,
$T_{gi}$ is the time constant of the governor, $T_{ti}$ is the time constant of the turbine,
$K_{pi}$ is the electric system gain, $T_{pi}$ is the electric system time constant,
$T_{12}$ is the tie-line synchronizing coefficient, and $r_i$ is the speed drop.

In Fig.~\ref{Fig6}, the system state $x\left( t \right)$ has nine elements. We need nine sensors,
each observing one dimension of $x\left( t \right)$ directly, and after sampling the measurement
signals are immediately sent to the LFC controllers.
Then, discretizing the process at the sampling instant gives the same formulas as
in (\ref{eq:2}) and (\ref{eq:3}) choosing $p = 2$.

In the simulation, based on \cite{IRef21}\cite{IRef22}, the sampling period is set to be
$h$=0.01, $T_{12}  = 2.4$, $K_{pi}  = 1,\ T_{pi}  = 0.2,\ T_{ti}  = 0.3,\ T_{gi}  = 0.08,\
r_i  = 0.2545,\ R_i = 1,\ i = 1,\ 2$, and
\begin{equation}
Q_{i,N}  = Q_i  = \left[ {\begin{array}{*{20}c}
   0 & 0 & 0 & 0 & 0 & 0 & 0 & 0 & 0  \\
   0 & 0 & 0 & 0 & 0 & 0 & 0 & 0 & 0  \\
   0 & 0 & 0 & 0 & 0 & 0 & 0 & 0 & 0  \\
   0 & 0 & 0 & 0 & 0 & 0 & 0 & 0 & 0  \\
   0 & 0 & 0 & 0 & 0 & 0 & 0 & 0 & 0  \\
   0 & 0 & 0 & 0 & 0 & 0 & 0 & 0 & 0  \\
   0 & 0 & 0 & 0 & 0 & 0 & 1 & 0 & 0  \\
   0 & 0 & 0 & 0 & 0 & 0 & 0 & 0 & 0  \\
   0 & 0 & 0 & 0 & 0 & 0 & 0 & 0 & 0  \\
\end{array}} \right],\ i = 1,\ 2.
\end{equation}

\begin{figure}[h]
  \begin{center}
  \includegraphics[width=0.9\textwidth]{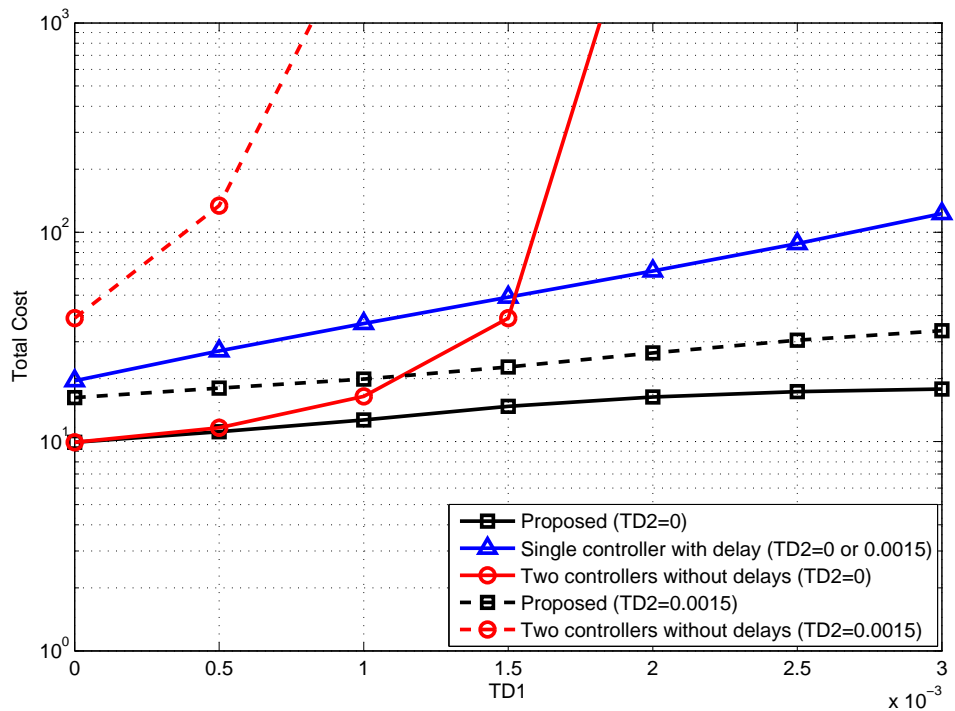}
  \caption{Performance comparison for three schemes under various TD1 for a two-area LFC system.} \label{Fig7}
  \end{center}
\end{figure}

In Fig.~\ref{Fig7}, we also compare the system performances of the three schemes.
Again we see that the proposed optimal distributed control strategy significantly outperforms
the other two methods.

\section{Conclusions}
We have considered the problem of optimal control for networked control systems
with distributed controllers and time delays under the linear quadratic control framework.
In particular, the optimal control problem is formulated as a non-cooperative linear quadratic game,
and we have obtained the optimal distributed controllers assuming the time delays between
sensors and actuators are deterministic and within one sampling period.
We have also applied the proposed optimal distributed
control scheme to load frequency control in power grid systems.
Future works include to investigate the optimal distributed
controller when the time delays are larger than one sampling period and stochastic.

\appendices
\section{}

From (\ref{eq:35}) and (\ref{eq:36}), we can rewrite $L_1 \left( k \right)$ as
\renewcommand{\theequation}{A-1}
\begin{equation}\label{eq:A1}
\begin{split}
 L_1 \left( k \right) &= \left( {P_{2,2}^1 \left( k \right)} \right)^{ - 1} P_{1,2}^1 \left( k \right) \\
 {\rm{        }} &= \left[ {\begin{array}{*{20}c}
   {a_1^1 \left( k \right) + a_2^1 \left( k \right)A_2 \left( k \right)} & {b_1^1 \left( k \right) + b_2^1 \left( k \right)B_{1,1}^2 \left( k \right)} & {c_1^1 \left( k \right) + c_2^1 \left( k \right)B_{2,1}^2 \left( k \right)}  \\
\end{array}} \right] ,\\
 \end{split}
\end{equation}
where
\renewcommand{\theequation}{A-2}
\begin{equation}\label{eq:A2}
\begin{split}
 a_1^1 \left( k \right) &= E_1^{ - 1} \left[ {\Gamma _{1,0}^T S_{1,1}^1 \left( {k + 1} \right)\Phi  + S_{2,1}^1 \left( {k + 1} \right)\Phi } \right]{\rm{ }} ,\\
 b_1^1 \left( k \right) &= E_1^{ - 1} \left[ {\Gamma _{1,0}^T S_{1,1}^1 \left( {k + 1} \right)\Gamma _{1,1}  + S_{2,1}^1 \left( {k + 1} \right)\Gamma _{1,1} } \right] ,\\
 c_1^1 \left( k \right) &= E_1^{ - 1} \left[ {\Gamma _{1,0}^T S_{1,1}^1 \left( {k + 1} \right)\Gamma _{2,1}  + S_{2,1}^1 \left( {k + 1} \right)\Gamma _{2,1} } \right] ,\\
 a_2^1 \left( k \right) &= b_2^1 \left( k \right) = c_2^1 \left( k \right) \\
 {\rm{        }} &= E_1^{ - 1} \left[ {\Gamma _{1,0}^T S_{1,1}^1 \left( {k + 1} \right)\Gamma _{2,0}  + S_{2,1}^1 \left( {k + 1} \right)\Gamma _{2,0}  + \Gamma _{1,0}^T S_{1,3}^1 \left( {k + 1} \right) + S_{2,3}^1 \left( {k + 1} \right)} \right] ,\\
 E_1  &= D_1^T S^1 \left( {k + 1} \right)D_1  + R_1  ,\\
 \end{split}
\end{equation}
and $S_{m,n}^1 \left( {k + 1} \right)$ is the $\left( {m,n} \right)$-$th$ block of
matrix $S^1 \left( {k + 1} \right)$, whose size is
$M \times M$ when $m = n = 1$, $M \times N$ when $m = 1$; $n \ge 2$, $N \times M$ when $m \ge 2$; $n = 1$,
and $N \times N$ when $m \ge 2$; $n \ge 2$.

From (A-1) and (\ref{eq:37}), we have
\renewcommand{\theequation}{A-3}
\begin{equation}\label{eq:A3}
\begin{split}
  - A_1 \left( k \right) &= a_1^1 \left( k \right) + a_2^1 \left( k \right)A_2 \left( k \right), \\
  - B_{1,1}^1 \left( k \right) &= b_1^1 \left( k \right) + b_2^1 \left( k \right)B_{1,1}^2 \left( k \right), \\
  - B_{2,1}^1 \left( k \right) &= c_1^1 \left( k \right) + c_2^1 \left( k \right)B_{2,1}^2 \left( k \right). \\
 \end{split}
\end{equation}

Similarly, we have
\renewcommand{\theequation}{A-4}
\begin{equation}\label{eq:A4}
\begin{split}
  - A_2 \left( k \right) &= a_1^2 \left( k \right) + a_2^2 \left( k \right)A_1 \left( k \right), \\
  - B_{1,1}^2 \left( k \right) &= b_1^2 \left( k \right) + b_2^2 \left( k \right)B_{1,1}^1 \left( k \right), \\
  - B_{2,1}^2 \left( k \right) &= c_1^2 \left( k \right) + c_2^2 \left( k \right)B_{2,1}^1 \left( k \right). \\
 \end{split}
\end{equation}
where
\renewcommand{\theequation}{A-5}
\begin{equation}\label{eq:A5}
\begin{split}
 a_1^2 \left( k \right) &= E_2^{ - 1} \left[ {\Gamma _{2,0}^T S_{1,1}^2 \left( {k + 1} \right)\Phi  + S_{3,1}^2 \left( {k + 1} \right)\Phi } \right], \\
 b_1^2 \left( k \right) &= E_2^{ - 1} \left[ {\Gamma _{2,0}^T S_{1,1}^2 \left( {k + 1} \right)\Gamma _{1,1}  + S_{3,1}^2 \left( {k + 1} \right)\Gamma _{1,1} } \right], \\
 c_1^2 \left( k \right) &= E_2^{ - 1} \left[ {\Gamma _{2,0}^T S_{1,1}^2 \left( {k + 1} \right)\Gamma _{2,1}  + S_{3,1}^2 \left( {k + 1} \right)\Gamma _{2,1} } \right], \\
 a_2^2 \left( k \right) &= b_2^2 \left( k \right) = c_2^2 \left( k \right) \\
 {\rm{        }} &= E_2^{ - 1} \left[ {\Gamma _{2,0}^T S_{1,1}^2 \left( {k + 1} \right)\Gamma _{1,0}  + S_{3,1}^2 \left( {k + 1} \right)\Gamma _{1,0}  + \Gamma _{2,0}^T S_{1,2}^2 \left( {k + 1} \right) + S_{3,2}^2 \left( {k + 1} \right)} \right], \\
 E_2  &= D_2^T S^2 \left( {k + 1} \right)D_2  + R_2 , \\
 \end{split}
\end{equation}
and $S_{m,n}^2 \left( {k + 1} \right)$ is the $\left( {m,n} \right)$-$th$ block of
matrix $S^2 \left( {k + 1} \right)$.

Then, based on (A-3) and (A-4), we can derive
\renewcommand{\theequation}{A-6}
\begin{equation}\label{eq:A6}
\begin{split}
 A_1 \left( k \right) &= \left[ {I - a_2^1 \left( k \right)a_2^2 \left( k \right)} \right]^{ - 1} \left[ {a_2^1 \left( k \right)a_1^2 \left( k \right) - a_1^1 \left( k \right)} \right], \\
 B_{1,1}^1 \left( k \right) &= \left[ {I - b_2^1 \left( k \right)b_2^2 \left( k \right)} \right]^{ - 1} \left[ {b_2^1 \left( k \right)b_1^2 \left( k \right) - b_1^1 \left( k \right)} \right], \\
 B_{2,1}^1 \left( k \right) &= \left[ {I - c_2^1 \left( k \right)c_2^2 \left( k \right)} \right]^{ - 1} \left[ {c_2^1 \left( k \right)c_1^2 \left( k \right) - c_1^1 \left( k \right)} \right], \\
 A_2 \left( k \right) &= \left[ {I - a_2^2 \left( k \right)a_2^1 \left( k \right)} \right]^{ - 1} \left[ {a_2^2 \left( k \right)a_1^1 \left( k \right) - a_1^2 \left( k \right)} \right], \\
 B_{1,1}^2 \left( k \right) &= \left[ {I - b_2^2 \left( k \right)b_2^1 \left( k \right)} \right]^{ - 1} \left[ {b_2^2 \left( k \right)b_1^1 \left( k \right) - b_1^2 \left( k \right)} \right], \\
 B_{2,1}^2 \left( k \right) &= \left[ {I - c_2^2 \left( k \right)c_2^1 \left( k \right)} \right]^{ - 1} \left[ {c_2^2 \left( k \right)c_1^1 \left( k \right) - c_1^2 \left( k \right)} \right]. \\
 \end{split}
\end{equation}

\ifCLASSOPTIONcaptionsoff
  \newpage
\fi

\newpage

\end{document}